# Efficient quantum teleportation of unknown qubit based on DV-CV interaction mechanism

**Sergey A. Podoshvedov**

Department of computer modeling and nanotechnology, Institute of natural and exact sciences, South Ural State University, Lenin Av. 76, Chelyabinsk, Russia
E-Mail: sapodo68@gmail.com

**Abstract:** We propose and develop theory of quantum teleportation of unknown qubit based on interaction mechanism between discrete-variable (DV) and continuous-variable (CV) states on highly transmissive beam splitter (HTBS). This DV-CV interaction mechanism is based on simultaneous displacement of the discrete-variable state on equal in absolute value but opposite in sign displacement amplitudes by coherent components of the hybrid, in such a way that all the information about the displacement amplitudes is lost with subsequent registration of photons in the auxiliary modes. The relative phase of the displaced unknown qubit in the measurement number state basis can vary on opposite depending on the parity of the basis states in the case of the negative amplitude of displacement that is akin to action of nonlinear effect on the teleported qubit. All measurement outcomes of the quantum teleportation are distinguishable, but the teleported state at Bob's disposal may acquire a predetermined amplitude-distorting factor. Two methods of getting rid of the factors are considered. The quantum teleportation is considered in various interpretations. Method for increasing the efficiency of quantum teleportation of an unknown qubit is proposed.

## 1. Introduction

Quantum nonlocality is a property of the universe that is independent of our description of nature. Quantum mechanical predictions on entangled quantum states cannot be simulated by any local hidden variable theory [1] that is confirmed in the experiments [2,3]. Bell's theorem [1] rules out local hidden local hidden variables to explain observed results. Although, in general case, quantum nonlocality is not equivalent to notion of entanglement, pure bipartite quantum state can most obviously manifest its nonlocal correlations. An example of the manifestation of the nonlocal nature of quantum objects is quantum teleportation [4]. Quantum entangled state connecting the sender and receiver of quantum information is used. In the protocol, an unknown quantum state of a physical system and a part of entangled state is measured in base of some states and subsequently reconstructed at a remote location (the physical components of the original system remain at the sending location) due to nonlocal nature of quantum channel. Quantum nonlocality does not allow for faster-than-communication [5], and hence is compatible with special relativity. Quantum teleportation can be reviewed as a protocol that most clearly demonstrates nonlocal trait of quantum entanglement. Quantum teleportation protocol is of interest as a conceptual as well as a basis for many other quantum protocols. Quantum teleportation protocol is used in schemes with quantum repeaters [6] being main ingredient for quantum communication over large distances. Quantum teleportation protocol underlies quantum gate teleportation [7], measurement-based computing [8]. The quantum teleportation protocol has been demonstrated in experiments using different physical systems and technologies. So, the quantum teleportation with polarization qubits was shown in [9]. Teleportation of unknown qubits of various nature through two-mode squeezed vacuum was demonstrated in [10-12]. Also, quantum teleportation has been achieved in laboratories including nuclear magnetic resonance [13], atomic ensembles [14], trapped atoms [15] and solid state systems [16].

Traditionally, when we talk about quantum teleportation, we mean quantum teleportation for two-level system called qubit [4]. Alice performs a joint quantum measurement, called Bell detection, which projects her unknown qubit and half a quantum channel into one of the states $(\sigma_i \otimes I)|\Psi\rangle$, where $\sigma_i$ is Pauli operator, $I$ is identical operator, $|\Psi\rangle$ is one of the four Bell states, $i = 0, \dots, 3$ and symbol $\otimes$ means tensor product. Alice's state of an unknown qubit disappears at her disposal, but in return, Bob simultaneously receives a state $\sum_{k=0}^{3} \sigma_i^+ \varrho \sigma_i$, where $\varrho$ is teleported qubit and $\sigma_i^+$ means Hermitian adjoint Pauli operator. Alice must communicate her measurement outcome $k$ to Bob, who then applies $\sigma_i$ and recovers the original unknown qubit $\varrho$. Despite its mathematical simplicity, the implementation of the complete Bell-states measurement faces a fundamental limitation [17]. Only two Bell states can be distinguished by linear optics methods, that limits the probability of success of quantum teleportation and the implementation of a controlled−$X$ gate by 0.5 and 0.25 [7], respectively. Attempts to circumvent this limitation are hardly possible due to the increasing difficulties in implementation [18-20]. So, multiparticle quantum entangled channel, which can hardly be generated in practice, with the subsequent registration of measurement outcomes exceeding 2 bits of classical information is required for teleportation of an unknown qubit with the success probability approaching unity in case of a significant increase of the number of the particles [18].

Quantum teleportation can also be extended to transmit information about quantum systems living in infinite-dimensional Hilbert space, known as continuous-variable (CV) systems. Vaidman has proposed teleportation of state of one-dimensional particle and CV quantum system using EPR-Borm pair [21]. Later, this idea was developed in representation of position- and momentum-like quadrature operators [22], now known as CV teleportation. CV teleportation can be made in a deterministic manner but with limited fidelity, in contrast to discrete-variable (DV) teleportation with fidelity of output state equal to one in ideal conditions. CV teleportation is applicable to transmitting both CV [10,11] and DV [12] states. Details of the CV states including CV quantum teleportation can be found in [23].

It was shown in [24] one cannot perform complete Bell-states measurement without "quantum-quantum" interaction which implies consideration of a hybrid physical system consisting of different ingredients, for example, atom and electromagnetic field in cavities. In general, a hybrid system may consist of components that may differ in nature, in size or in their description. So, in the case of using light, we can consider hybrid systems that are formed by DV and CV states [25]. Recently, the possibility of generating [26] and manipulating [27] hybrid entangled states has been shown. The hybrid entangled states that are formed from number states and their displaced analogues or the same displaced number states [28-30], are of interest. The implementation of the displaced states of light has been discussed in [31,32]. Here we offer a new type of quantum teleportation of unknown qubit which is based on nonlinear effect of interaction of DV and CV states on HTBS. Such an approach aims to make use of advantages of DV and CV states to teleport unknown qubit with larger success probability and high fidelity. The proposed approach differs from DV and CV teleportation, but can be recognized as being closer to CV one. Hybrid entanglement hybrid entanglement formed by coherent components with different in sign amplitudes and dual-single photon is used for transmission of quantum information from sender to receiver. Non-linear effect on the target state in Bob's hands is realized due to interaction of CV and DV states on HTBS [26,33,34] (DV-CV interaction mechanism). Various interpretations of the DV-CV quantum teleportation of an unknown qubit are reviewed and found, to date, the best strategies for increasing its efficiency in terms of success probability.

## 2. DV-CV quantum teleportation of unknown qubit via hybrid non-maximally entangled state

Consider the following hybrid entangled state as quantum channel for the quantum teleportation of unknown qubit

$$|\Psi\rangle_{123} = (|-\beta\rangle_1|01\rangle_{23} + |\beta\rangle_1|10\rangle_{23})/\sqrt{2}, \qquad (1)$$

where the subscript denotes the number of the mode. The hybrid entangled state consists of the coherent components with opposite in sign amplitudes (here and in the following the amplitude is assumed to be positive $\beta > 0$) and the single photon taking simultaneously two mode (dual-rail single photon). The state (1) is non-maximally entangled state due to non-orthogonality of the coherent states. Negativity which is easy to compute in four-dimensional Hilbert space can be taken as a measure of the quantum entanglement [35]. The quantity is derived from PPT criterion for separability [36] and possesses all proper properties for the entanglement measure. The negativity of composed system can be defined in terms of the density matrix $\varrho$ as $\tau = (\|\varrho^{T_A}\| - 1)/2$, where $\varrho^{T_A}$ is the partial transpose of $\varrho$ with respect to subsystem $A$ of two-partite system $AB$ and $\|\varrho^{T_A}\| = tr|\varrho^{T_A}| = tr\sqrt{(\varrho^{T_A})^+\varrho^{T_A}}$ is the trace norm of the sum of the singular values of the operator $\varrho^{T_A}$, where $(\varrho^{T_A})^+$ means Hermitian conjugate operator of original $\varrho^{T_A}$. The negativity takes the maximum value $\tau_{max} = 1$ for maximally entangled states. Doing the calculations for the state (1), one obtains

$$\tau = \sqrt{1 - exp(-4|\beta|^2)}. \qquad (2)$$

The negativity of the hybrid state (1) attains maximal value $\tau \to \tau_{max}$ in the case of an infinitely large value of the amplitude of the coherent states $\beta \to \infty$. Otherwise, the hybrid state (1) is non-maximally entangled state. Although for sufficiently large values of the amplitude $\beta$ of the coherent states, the hybrid state (1) can be considered as almost maximal one $\tau \approx \tau_{max}$ since the exponential factor decreases rapidly enough.

Now, we are going to use non-maximally entangled state (1) to teleport unknown qubit, in general case, represented by the following superposition

$$|\varphi^{(lk)}\rangle_{12} = a_0|lk\rangle_{12} + a_1|kl\rangle_{12}, \qquad (3)$$

satisfying the normalization condition $|a_0|^2 + |a_1|^2 = 1$ with qubit's amplitudes $a_0$ and $a_1$ unknown to anyone, where $|l\rangle$ and $|k\rangle$ are the arbitrary number (Fock) states. In particular, we have unknown dual-rail single photon

$$|\varphi^{(01)}\rangle_{12} = a_0|01\rangle_{12} + a_1|10\rangle_{12}, \qquad (4)$$

in the case of $l = 0$ and $k = 1$. Consider the optical scheme in Fig. 1 adjusted for teleportation of the unknown qubit. Alice and Bob are the standard participants of the protocol who can be at considerable distance from each other. The hybrid entangled state $|\Psi\rangle_{156}$ (Eq. (1)) in modes 1, 5 and 6 is used as quantum channel for the quantum teleportation, where coherent part in mode 1 belongs to Alice while single photon taking simultaneously both fifth and sixth modes is in Bob's location. Unknown qubit $|\varphi^{(lk)}\rangle_{34}$ is at the disposal of Alice. In addition to the states, Alice uses ancillary coherent state with real amplitude $\beta_1 > 0$ $|-\beta_1\rangle_2$ taking the second mode to mix it with one of modes of the unknown qubit of beam splitter, where, in general case $\beta_1 \neq \beta$. The optical scheme in Fig. 1 operates in linear optics domain with optical elements and photodetectors. Key moment of the quantum teleportation implementation is to provide discrete-continuous interaction between coherent components and unknown qubit. The discrete-continuous interaction is realized on highly transmissive beam splitter which is described by the following unitary matrix

$$BS_{13} = \begin{bmatrix} t & -r \\ r & t \end{bmatrix}, \qquad (5a)$$

where the real parameters $t > 0$, $r > 0$ are the transmittance $t \to 1$ and reflectance $r \to 0$, respectively, satisfying the normalization condition $t^2 + r^2 = 1$. Here, subscripts 13 imply the first mode of the state (1) and third mode of the unknown qubit (3) are mixed on the HTBS. Another HTBS

$$BS_{24} = \begin{bmatrix} t_1 & -r_1 \\ r_1 & t_1 \end{bmatrix}, \tag{5b}$$

is used to mix ancillary coherent state with fourth mode of the teleported qubit (the subscript 24 is used in (5b) to discriminate the beam splitter from one (5a)). Here, the real beam splitter parameters obey the similar condition $t_1^2 + r_1^2 = 1$ and, in general case, $t_1 \neq t$ and $r_1 \neq r$. Interaction of discrete- and continuous-variable states ends in measurements performed in the modes 1, 3 and 4 leaving the state in mode 2 untouched to collapse Bob's state into a new in dependence on Alice's measurement outcomes. All information about the teleported qubit disappears in measurement process. Alice can communicate with Bob with negligible number of the classical information to help him to recover the original state.

Strong coherent pumping $|\beta\rangle$ displaces an arbitrary state $\rho$ by some amount, provided that the beam splitter transmits a significant part of the input light $t \to 1$ [37]

$$BS(\rho \otimes |\beta\rangle\langle\beta|)BS^+ \approx D(\alpha)\rho D^+(\alpha) \otimes |\beta\rangle\langle\beta|, \tag{6}$$

where the displacement operator $D(\alpha)$ [37] with displacement amplitude $\alpha$ is used, symbol $\otimes$ means tensor product of two operators and $D^+(\alpha)$ is Hermitian conjugate of the operator $D(\alpha)$. The amplitude of the displacement is given by

$$\alpha = \beta r/t \approx \beta r, \tag{7}$$

in the case of $t \approx 1$. The same reasoning is applicable to interaction of arbitrary state $\rho$ with the coherent state $|-\beta\rangle$ with output approximate state

$$BS(\rho \otimes |-\beta\rangle\langle-\beta|)BS^+ \approx D(-\alpha)\rho D^+(-\alpha) \otimes |-\beta\rangle\langle-\beta|. \tag{8}$$

Note the condition (7) means that amplitude of the coherent state must tend to infinity $\beta \to \infty$ if $r \to 0$ to keep exact condition $\alpha = \beta r = const$. But in real experiment with the non-zero reflectance $r \neq 0$, the amplitude of the coherent states takes large but nevertheless finite values sufficient to satisfy the condition (7). For this reason, approximate equality is used in Eqs. (6) and (8) which goes into the exact equality in the limit case of $t \to 1$. The better we fulfill the condition $r \to 0$ and $\beta \to \infty$, with higher fidelity the output states are close to ideal ones on the right-hand side of the Eqs. (6) and (8).

Now, we are going to make use of mathematical apparatus developed in [28,33,34] with displaced number states defined with help of the displacement operator $|n, \alpha\rangle = D(\alpha)|n\rangle$ [28]. The states are orthogonal $\langle n, \alpha | m, \alpha \rangle = \delta_{nm}$ with $\delta_{nm}$ being Kronecker delta [28]. The displaced number states are defined by two numbers: quantum discrete number $n$ and classical continuous parameter $\alpha$ which can be recognized as their size. The partial case is the infinite set of the number states $\{|n\rangle, n = 0,1,2, \ldots, \infty\}$ with $\alpha = 0$. Here we are going to make use of the completeness of the Fock's states $\sum_{n=0}^{\infty} |n\rangle\langle n| = I$ to decompose arbitrary displaced number states state $|l, \alpha\rangle$ over them [34]

$$|l, \alpha\rangle = F \sum_{n=0}^{\infty} c_{ln}(\alpha)|n\rangle, \tag{9}$$

where the overall multiplier $F(\alpha) = exp(-|\alpha|^2/2)$ is introduced. The matrix elements $c_{ln}(\alpha)$ satisfy the normalization condition $F^2 \sum_{n=0}^{\infty} |c_{ln}(\alpha)|^2 = 1$ [34]. In particular, the matrix elements $c_{0n}(\alpha) = \alpha^n/\sqrt{n!}$ are the amplitudes of the coherent state $|\alpha\rangle \equiv |0, \alpha\rangle$ [28]. All other matrix elements with $l \neq 0$ are presented in [34].

The realization of the nonlinear effect in DC interaction is ensured by the property of matrix elements to change their sign when changing the displacement amplitude on opposite in sing $\alpha \to -\alpha$. The matrix elements change as

$$c_{ln}(-\alpha) = (-1)^{n-l} c_{ln}(\alpha). \tag{10}$$

under change of the displacement amplitude on opposite $\alpha \to -\alpha$ [34]. In particular, we have the following relation for the matrix elements of even $l = 2m$ displaced number states

$$c_{2mn}(-\alpha) = (-1)^n c_{2mn}(\alpha), \tag{11}$$

and for the matrix elements of odd $l = 2m + 1$ displaced number states and
$$c_{2m+1n}(-\alpha) = (-1)^{n-1}c_{2m+1n}(\alpha), \qquad (12)$$
for the decomposition of odd $l = 2m + 1$ displaced number states. In particular, we have $c_{0n}(-\alpha) = (-1)^n c_{0n}(\alpha)$ for the amplitudes of the coherent state. This difference in the behavior of the matrix elements when changing parity of the displaced number states (Eqs. (11,12)) is similar to a nonlinear action of two-qubit gate controlled−Z gate. Coherent components of the hybrid entangled state (1) simultaneously displace the unknown teleported qubit (3) in indistinguishable manner on HTBS as given by Eqs. (6,8), respectively, by the values that differ from each other only by sign. All information about value of the displacement of the teleported qubit (either by $\alpha$ or $-\alpha$) disappears. Measurement of the unknown teleported state and coherent part of the state (1) collapses the original state $BS_{13}BS_{24}\left(|\Psi\rangle_{156}|0,-\beta_1\rangle_2|\varphi^{(lk)}\rangle_{34}\right)$ onto state at Bob's disposal subject controlled−Z operation in the case of corresponding parity of the number states $|l\rangle$ and $|k\rangle$ in (3) and the teleported state can be recovered through classical communication.

Let us present mathematical details of interaction of hybrid non-maximally entangled state (1) and ancillary coherent state with unknown qubit on two HTBS (5a) and (5b) as shown in Fig. 1. Due to linearity of the beam splitter operators, we have
$$BS_{13}BS_{24}\left(|\Psi\rangle_{156}|0,-\beta_1\rangle_2|\varphi^{(lk)}\rangle_{34}\right) = (1/\sqrt{2})$$
$$\left(BS_{13}BS_{24}\left(|0,-\beta\rangle_1|0,-\beta\rangle_1|\varphi^{(lk)}\rangle_{34}\right) + BS_{13}BS_{24}\left(|0,-\beta\rangle_1|0,-\beta\rangle_1|\varphi^{(lk)}\rangle_{34}\right)\right), \quad (13)$$
where the hybrid non-maximally entangled state (1) is considered to take modes 1, 5 and 6, the teleported unknown qubit is located in modes 3 and 4, while ancillary coherent state is used in second mode. Consider action of the beam splitters on the states separately. Then, we have [38]
$$BS_{13}BS_{24}\left(|0,-\beta\rangle_1|0,-\beta\rangle_2|\varphi^{(lk)}\rangle_{34}\right)|01\rangle_{56} =$$
$$BS_{13}BS_{24}(|0,-\beta\rangle_1|0,-\beta\rangle_2(a_0|01\rangle_{34} + a_1|10\rangle_{34}))|01\rangle_{56} \rightarrow$$
$$F^2|0,-\beta/t\rangle_1|0,-\beta/t_1\rangle_2 \sum_{n=0}^{\infty}\sum_{m=0}^{\infty} t^{n+m} c_{ln}(\alpha)c_{km}(\alpha_1)$$
$$\left(a_0 + a_1 A_{nm}^{(lk)}\right)|nm\rangle_{34}|01\rangle_{56} +$$
$$rF^2\begin{pmatrix}\sum_{n=0}^{\infty}\sum_{m=0}^{\infty} t^{n+m-1} c_{ln}(\alpha)c_{km}(\alpha_1)\left(a_0 + a_1 A_{nm}^{(lk)}\right) \\ \left(\sqrt{n}|1,-\beta/t\rangle_1|0,-\beta/t_1\rangle_2|n-1m\rangle_{34} + \right. \\ \left. \sqrt{m}|0,-\beta/t\rangle_1|1,-\beta/t_1\rangle_2|nm-1\rangle_{34}\right)\end{pmatrix}|01\rangle_{56}, \qquad (14)$$
for the first term in Eq. (13) and
$$BS_{13}BS_{24}\left(|0,\beta\rangle_1|0,-\beta\rangle_2|\varphi^{(lk)}\rangle_{34}\right)|10\rangle_{56} =$$
$$BS_{13}BS_{24}(|0,\beta\rangle_1|0,-\beta\rangle_2(a_0|01\rangle_{34} + a_1|10\rangle_{34}))|10\rangle_{56} \rightarrow$$
$$F^2|0,\beta/t\rangle_1|0,-\beta/t_1\rangle_2 \sum_{n=0}^{\infty}\sum_{m=0}^{\infty} (-1)^{n-l} t^{n+m} c_{ln}(\alpha)c_{km}(\alpha_1)$$
$$\left(a_0 + (-1)^{l-k} a_1 A_{nm}^{(lk)}\right)|nm\rangle_{34}|10\rangle_{56}$$
$$+rF^2\begin{pmatrix}\sum_{n=0}^{\infty}\sum_{m=0}^{\infty}(-1)^{n-l} t^{n+m-1} c_{ln}(\alpha)c_{km}(\alpha_1)\left(a_0 + (-1)^{l-k} a_1 A_{nm}^{(lk)}\right) \\ \left(\sqrt{n}|1,\beta/t\rangle_1|0,-\beta/t_1\rangle_2|n-1m\rangle_{34} + \right. \\ \left. \sqrt{m}|0,\beta/t\rangle_1|1,-\beta/t_1\rangle_2|nm-1\rangle_{34}\right)\end{pmatrix}|10\rangle_{56}, \quad (15)$$
for the second term in Eq. (13), where amplitude-distorting coefficients $A_{nm}^{(lk)}$ are given by
$$A_{nm}^{(lk)}(\alpha,\alpha_1) = \frac{c_{kn}(\alpha)c_{lm}(\alpha_1)}{c_{ln}(\alpha)c_{km}(\alpha_1)}. \qquad (16)$$
Note the displacement amplitude $\alpha_1$ is determined by $\alpha_1 = \beta_1 r/t \approx \beta_1 r$ (Eq. (7)). Here, we limited ourselves by the first two terms in order of smallness $r \ll 1$ neglecting members of higher order of smallness in the reflectance proportional to $\sim r^n$ with $n > 1$. First terms of zeroth order in $r$ give maximal contribution, while influence of the second terms proportional to $\sim r$ goes to zero in the case of $r \to 0$.

Consider output state in ideal case of $t = 1$ and $r = 0$ in terms of even/odd superposition of coherent states (SCS) defined by

$$|even\rangle = N_+(|-\beta\rangle + |\beta\rangle), \tag{17}$$
$$|odd\rangle = N_-(|-\beta\rangle - |\beta\rangle), \tag{18}$$

where the factors $N_\pm = \left(2(1 \pm exp(-2|\beta|^2))\right)^{-1/2}$ are the normalization parameters. Then, we can approximate the state $BS_{13}BS_{24}\left(|\Psi\rangle_{156}|0,-\beta_1\rangle_2|\varphi^{(lk)}\rangle_{34}\right)$ in zeroth order on parameter $r \ll 1$

$$BS_{13}BS_{24}\left(|\Psi\rangle_{156}|0,-\beta_1\rangle_2|\varphi^{(lk)}\rangle_{34}\right) \approx (F^2/2)|0,-\beta_1\rangle_2$$
$$\sum_{n=0}^\infty \sum_{m=0}^\infty c_{ln}(\alpha)c_{km}(\alpha_1)N_{nm}^{(lk)-1}\left(\frac{|even\rangle_1}{N_+}|\Psi_{nm}^{(lk)}\rangle_{56} + \frac{|odd\rangle_1}{N_-}|\Psi_{n+1m}^{(lk)}\rangle_{56}\right)|nm\rangle_{34}, \tag{19}$$

where the state at Bob's location (Bob's states) becomes

$$|\Psi_{nm}^{(lk)}\rangle_{56} = \frac{N_{nm}^{(lk)}}{\sqrt{2}}\left(\left(a_0 + a_1 A_{nm}^{(lk)}\right)|01\rangle_{56} + (-1)^{n-l}\left(a_0 + (-1)^{l-k}a_1 A_{nm}^{(lk)}\right)|10\rangle_{56}\right), \tag{20}$$

$$|\Psi_{n+1m}^{(lk)}\rangle_{56} = \frac{N_{nm}^{(lk)}}{\sqrt{2}}\left(\left(a_0 + a_1 A_{nm}^{(lk)}\right)|01\rangle_{56} + (-1)^{n+1-l}\left(a_0 + (-1)^{l-k}a_1 A_{nm}^{(lk)}\right)|10\rangle_{56}\right), \tag{21}$$

where the normalization factor $N_{nm}^{(lk)}$ is given by

$$N_{nm}^{(lk)} = \left(|a_0|^2 + |a_1|^2\left|A_{nm}^{(lk)}\right|^2\right)^{-1/2} = \left(1 + \left(\left|A_{nm}^{(lk)}\right|^2 - 1\right)|a_1|^2\right)^{-1/2}. \tag{22}$$

To provide the performance of nonlinear action of controlled$-Z$ gate

$$(-1)^{l-k} = -1, \tag{23}$$

we need to impose additional requirement on the teleported qubit (3), namely, difference $l - k$ must be odd number for used displacement amplitude $\beta > 0$ of the hybrid non-maximally entangled state (1). For example, if we take $l = 0$ and $k = 1$ (dual-rail unknown single photon), we provide performance of the condition (23).

Now, Alice must do the parity measurement at first mode to recognize even/odd SCS and registers the measurement outcome $|nm\rangle_{34}$ in measured third and fourth modes. Then, Bob obtains one of the two states either $|\Psi_{nm}^{(lk)}\rangle_{56}$ (Eq. (20)) or $|\Psi_{n+1m}^{(lk)}\rangle_{56}$ (Eq. (21)) in dependence on parity of the measured photons at mode 1. Assume that Alice registers only definite measurement outcome $(nm)$ and informs Bob about it. Then, Bob can apply sequence of operators of Hadamard gate and $Z-$ gate in some power to get

$$HZ^{n-l}|\Psi_{nm}^{(lk)}\rangle = N_{nm}^{(lk)}\begin{bmatrix}a_0\\a_1 A_{nm}^{(lk)}\end{bmatrix}, \tag{24}$$

$$HZ^{n-l+1}|\Psi_{n+1m}^{(lk)}\rangle = N_{nm}^{(lk)}\begin{bmatrix}a_0\\a_1 A_{nm}^{(lk)}\end{bmatrix}. \tag{25}$$

$Z-$ gate is applied in dependence on the parity of the numbers $n-l$ and $n-l+1$ as $Z^2 = I$, where $I$ is an identical operator. Hadamard operation is applied regardless of whether Bob should initially use $Z-$gate or not. These operations (Hadamard gate and $Z-$ gate) are easily implemented by linear optics devices on single photon [40]. Obtained state contains amplitude-distorting factor $A_{nm}^{(lk)}$ defined by Eq. (16). We are going to call such states as amplitude-modulated (AM) states. The presence of this additional factor $A_{nm}^{(lk)}$ is a distinctive feature inherent to DV-CV interaction. One can even say that the CV state leaves its imprint in the teleporting DV state. The success probability for Alice to register the measurement outcome $|nm\rangle_{34}$ not depending on parity of the states in first mode is given by

$$P_{nm}^{(lk)} = \frac{F^4|c_{ln}(\alpha)|^2|c_{km}(\alpha_1)|^2}{N_{nm}^{(lk)2}}, \tag{26}$$

where the probabilities are normalized
$$\sum_{n=0}^{\infty} \sum_{m=0}^{\infty} P_{nm}^{(lk)} = 1, \quad (27)$$
not depending on the numbers $l$ and $k$ that can be directly checked using normalization of the matrix elements $c_{nm}(\alpha)$. It is worth noting the success probabilities of the measurement outcomes $P_{nm}^{(lk)}$ depend on the displacement amplitudes $\alpha$ and $\alpha_1$ and can change in wide diapason. In other words, Alice has additional parameters which she manipulate to vary the success probabilities of her measurement outcomes.

Consider the case of $\alpha = \alpha_1$ that can be produced by application of coherent states with equal displacement amplitudes $\beta = \beta_1$ that displace the teleported qubit on equivalent HTBS (Eq. (5a) and (5b)). Then, by definition (16), we have
$$A_{nn}^{(lk)} = 1. \quad (28)$$
This means that the probabilistic protocol of the DV-CV quantum teleportation of an unknown qubit can be realized if Alice registers only the same measurement outcomes $n = m$ together with parity measurement at first mode by discarding all other $n \neq m$. Moreover, Alice must transmit one bit of classical information over the classical communication channel to indicate to Bob whether he should apply $Z-$ transformation in the probabilistic teleportation. The success probability of the event is equal to
$$P_T^{(lk)} = \sum_{n=0}^{\infty} P_{nn}^{(lk)} = F^4 \sum_{n=0}^{\infty} |c_{ln}(\alpha)|^2 |c_{kn}(\alpha)|^2. \quad (29)$$
In all remaining cases $n \neq m$, the Bob's qubit receives an additional amplitude-distorting factor $A_{nm}^{(lk)}$ not equal to one being a price for implementation of controlled$-Z$ operation in DV-CV interaction. But the factor is known to both participants of the protocol provided that they know the displacement amplitude $\alpha$ and measurement outcomes $n$ and $m$. The probability for Bob to receive AM qubit (after receiving relevant auxiliary classical information from Alice) is equal to
$$P_{AM}^{(lk)} = \sum_n^{\infty} \sum_{m,n \neq m}^{\infty} P_{nm}^{(lk)}. \quad (30)$$
Thus, the total probability can be divided into two categories: the success probability to perfectly teleport unknown qubit (29) only with one bit of assisting classical information and probability to transmit to Bob AM qubit with some amount of auxiliary classical information
$$P_T^{(lk)} + P_{AM}^{(lk)} = 1. \quad (31)$$
It is worth noting that both $P_{nm}^{(lk)}$ with $n \neq m$ (26) also depend (in addition to dependence on the displacement amplitude $\alpha$) on the parameters of the teleported qubit (3), namely on the amplitude $|a_1|$ due to the amplitude-distorting factor $A_{mn}^{(lk)}$ in the normalization multiplier $N_{nm}^{(lk)}$. When receiving AM qubits, Bob can take certain measures to get rid of the amplitude-distorting factors.

Note only the amplitude factor obey the condition
$$A_{mn}^{(lk)} = \left(A_{nm}^{(lk)}\right)^{-1}, \quad (32)$$
in the case of $\alpha = \alpha_1$. Using the relation, it is possible to show that sum of two probabilities $P_{nm}^{(lk)}$ and $P_{mn}^{(lk)}$ does not depend on the amplitude $|a_1|$ of the teleported unknown qubit
$$P_{nm}^{(lk)S} = P_{nm}^{(lk)} + P_{mn}^{(lk)} = F^4(|c_{ln}|^2|c_{km}|^2 + |c_{lm}|^2|c_{kn}|^2) =$$
$$F^4 |c_{ln}|^2 |c_{km}|^2 \left(1 + \left|A_{nn}^{(lk)}\right|^2\right), \quad (33)$$
where superscript $S$ concerns the sum of two probabilities. It proves the fact that the total probability $P_{AM}^{(lk)}$ (Eq. (30)) also does not depend on the parameter $|a_1|$ of the teleported qubit in spite of the fact that each member $P_{nm}^{(lk)}$ of this sum still depends on the parameters $|a_1|$ of the teleported qubit. Finally, the probability for Bob to obtain AM originally unknown qubit can be rewritten as
$$P_{AM}^{(lk)} = \sum_n^{\infty} \sum_{m,n \neq m}^{\infty} P_{nm}^{(lk)S} = F^4 \sum_n^{\infty} \sum_{m,n \neq m}^{\infty} (|c_{ln}|^2|c_{km}|^2 + |c_{lm}|^2|c_{kn}|^2). \quad (34)$$

The proposed method of implementing DV-CV quantum teleportation can also be used for the unknown single-rail unknown qubit composed of $|l\rangle$ and $|k\rangle$ photons

$$|\phi^{(lk)}\rangle_1 = a_0|l\rangle_1 + a_1|k\rangle_1. \tag{35}$$

In particular, the unknown single-rail qubit $|\phi^{(01)}\rangle$ is the superposition of vacuum and single photon. The same state in Eq. (1) is used as quantum channel for quantum teleportation of unknown qubit (35). In this case, we can also use the scheme in Figure 1, but only without interacting with the additional coherent state $|-\beta_1\rangle_2$. Then, following the same technique, we obtain

$$BS_{12}\left(|\Psi\rangle_{134}|\phi^{(lk)}\rangle_2\right) \to (F/2)$$
$$\sum_{n=0}^{\infty} c_{ln}(\alpha) N_n^{(lk)-1} \left(\frac{|even\rangle_1}{N_+}|\Psi_n^{(lk)}\rangle_{34} + \frac{|odd\rangle_1}{N_-}|\Psi_{n+1}^{(lk)}\rangle_{56}\right)|n\rangle_2, \tag{36}$$

in the case of $t \to 1$ and $r \to 0$. Another difference from the formula (19) is that the real amplitude-distorting factors $A_n^{(lk)}$ in the states $|\Psi_n^{(lk)}\rangle$ and $|\Psi_{n+1}^{(lk)}\rangle$ are determined by

$$A_n^{(lk)} = \frac{c_{kn}(\alpha)}{c_{ln}(\alpha)}, \tag{37}$$

where the states in Bob's location are the same as in Eqs. (20,21) with the normalization factors $N_n^{(lk)} = \left(1 + \left(\left|A_n^{(lk)}\right|^2 - 1\right)|a_1|^2\right)^{-1/2}$. If Alice performs the parity measurement in the first mode and determines the number of photons in the second measurement mode, then she collapses the initial state into one of the possible states either (20) or (21). Then, she can send Bob additional classic information so that he can make corresponding unitary transformations with his qubit to get the AM state with known factor $A_n^{(lk)}$

$$N_n^{(lk)} \begin{bmatrix} a_0 \\ a_1 A_n^{(lk)} \end{bmatrix}, \tag{38}$$

with success probability

$$P_n^{(lk)} = \frac{F^2 |c_{ln}(\alpha)|^2}{N_n^{(lk)2}}. \tag{39}$$

From the comparison of amplitude-distorting coefficients $A_{nm}^{(lk)}$ and $A_n^{(lk)}$, we can see a difference in the two types of DV-CV quantum teleportation of unknown qubit. Registration of identical outcomes $n = m$ in two auxiliary measurement modes leads to the fact that Bob's state gets rid of these additional parameters $A_{nn}^{(lk)}$ (Eq. (28)). Teleportation of the single-rail initial state (35) without amplitude-distorting parameter $A_n^{(lk)} = \pm 1$ is possible if $c_{kn}(\alpha) = \pm c_{ln}(\alpha)$.

## 3. Methods to increase the success probabilities of the DV-CV quantum teleportation

We showed in the previous part that the DV-CV quantum teleportation protocol allows us to transfer to Bob either the original unknown qubit or its amplitude-distorted version. All measurement outcomes give different states and all amplitude-distorting coefficients are known in advance. The implementation of the DV-CV protocol takes place in a deterministic manner, but the fidelity of the output state, in the general case, is not ideal equal to one. Therefore, our efforts are now focused on the opportunity for Bob to restore the initial state from AM qubit with help of communication with Alice. To consider methods to increase the success probabilities of the quantum teleportation, let us present matrix elements for first six displaced number states. So, we have for the coherent state $|0, \alpha\rangle$

$$c_{0n}(\alpha) = \alpha^n/\sqrt{n!}, \tag{40}$$

for the displaced singe photon $|1, \alpha\rangle$

$$c_{1n}(\alpha) = \alpha^{n-1}(n - |\alpha|^2)/\sqrt{n!}, \tag{41}$$

for the displaced two-photon state $|2, \alpha\rangle$

$$c_{2n}(\alpha) = \alpha^{n-2}(n(n-1) - 2n|\alpha|^2 + |\alpha|^4)/(\sqrt{2}\sqrt{n!}), \tag{42}$$

for the displaced three-photon state $|3,\alpha\rangle$

$$c_{3n}(\alpha) = \alpha^{n-3}(n(n-1)(n-2) - 3n(n-1)|\alpha|^2 + 3n|\alpha|^4 - |\alpha|^6)/(\sqrt{3!}\sqrt{n!}), \tag{43}$$

for the displaced four-photon state $|4,\alpha\rangle$

$$c_{4n}(\alpha) = \alpha^{n-4}\begin{pmatrix} n(n-1)(n-2)(n-3) - 4n(n-1)(n-2)|\alpha|^2 + \\ 6n(n-1)|\alpha|^4 - 4n|\alpha|^6 + |\alpha|^8 \end{pmatrix}/(\sqrt{4!}\sqrt{n!}), \tag{44}$$

for the displaced state with five photons $|5,\alpha\rangle$

$$c_{5n}(\alpha) = \alpha^{n-5}\begin{pmatrix} n(n-1)(n-2)(n-3)(n-4) - \\ 5n(n-1)(n-2)(n-3)|\alpha|^2 + \\ 10n(n-1)(n-2)|\alpha|^4 - 10n(n-1)|\alpha|^6 + \\ 5n|\alpha|^8 - |\alpha|^{10} \end{pmatrix}/(\sqrt{5!}\sqrt{n!}). \tag{45}$$

Using the expressions and formulas $A_{nm}^{(lk)}$ (Eq. (16)) and $A_n^{(lk)}$ (Eq. (37)), we can calculate any amplitude-distorting factor for any teleported unknown qubit.

Suppose that Bob can demodulate his AM unknown qubit either $N_{nm}^{(lk)}\left(a_0|01\rangle + a_0 A_{nm}^{(lk)}|01\rangle\right)$ or $N_n^{(lk)}\left(a_0|01\rangle + a_0 A_n^{(lk)}|01\rangle\right)$ with the probability $q_{nm}^{(lk)}$. Then, we get the next addition to the overall success probability of DV-CV quantum teleportation

$$\delta P_T^{(lk)} = F^4 \sum_n^\infty \sum_{m,n\neq m}^\infty \left( q_{nm}^{(lk)}|c_{ln}|^2|c_{km}|^2 + q_{mn}^{(lk)}|c_{lm}|^2|c_{kn}|^2 \right), \tag{46}$$

where the overall success probability $P_T^{(lk)O}$ becomes

$$P_T^{(lk)O} = P_T^{(lk)} + \delta P_T^{(lk)}. \tag{47}$$

Here, the normalization factor $N_{nm}^{(lk)}$ in expression for the success probability disappears as we get rid of the amplitude-distorting factor $A_{nm}^{(lk)}$. Similar addition to the success probability can be obtained in the case of amplitude demodulation of an unknown qubit $N_n^{(lk)}\left(a_0|01\rangle + a_0 A_n^{(lk)}|01\rangle\right)$.

Amplitude demodulation of an unknown qubit (or the same deliverance from amplitude-distorting factor) may not be an easy task. It seems that this operation could be performed at the next conversion: $|01\rangle \to |01\rangle$ and $|10\rangle \to \exp(\pm\Gamma)|10\rangle$, where either $\exp(\pm\Gamma) = A_{nm}^{(lk)-1}$ or $\exp(\pm\Gamma) = A_n^{(lk)-1}$ in dependency on $A_{nm}^{(lk)} < 1$, $A_n^{(lk)} < 1$ or $A_{nm}^{(lk)} > 1$, $A_n^{(lk)} > 1$ with $\Gamma$ being some either amplifying or weakening parameter. The conversion is not unitary. Consider more realistic scheme for amplitude demodulation of unknown qubit $N_{nm}^{(lk)}\left(a_0|01\rangle_{12} + a_1 A_{nm}^{(lk)}|10\rangle_{12}\right)$. Reconstruction of the original state [40] is probabilistic provided that some measurement outcome is fixed in auxiliary mode. The mode 2 in the state is auxiliary. The displacement operator $D_2(\gamma)$ with amplitude $\gamma$ acts on second mode of the state producing

$$D_2(\gamma) N_{nm}^{(lk)}\left(a_0|01\rangle_{12} + a_1 A_{nm}^{(lk)}|10\rangle_{12}\right) = $$
$$F(\gamma) \sum_{n=0}^\infty c_{1n}(\gamma)\left(a_0|0\rangle_1 + a_1 A_{nm}^{(lk)}(c_{0n}(\gamma)/c_{1n}(\gamma))|1\rangle_1\right)|n\rangle_2. \tag{48}$$

Measurement of the $|n\rangle$ photons in second mode generates the following state (leaving out normalization factor) $a_0|0\rangle_1 + a_1 A_{nm}^{(lk)}(c_{0n}(\gamma)/c_{1n}(\gamma))|1\rangle_1$ [40] which is converted into original one provided the following condition

$$A_{nm}^{(lk)}(\alpha)(c_{0n}(\gamma)/c_{1n}(\gamma)) = \pm 1, \tag{49}$$

is satisfied. Then, the success probability of the amplitude demodulation through the displacement operator is given by

$$q_{nm}^{(lk)D} = F^2(\gamma)|c_{1n}(\gamma)|^2, \tag{50}$$

where value of the parameter $\gamma$ follows from (49) and superscript $D$ means the original state is obtained with help of mixing it with coherent state.

Consider another way to get rid of amplitude factor $A_{nm}^{(lk)}$ in unknown qubit. To do it we are going to make use of quantum swapping method [41] when AM unknown qubit $N_{nm}^{(lk)}\left(a_0|01\rangle_{12} + a_1 A_{nm}^{(lk)}|10\rangle_{12}\right)$ interacts with the prearranged state

$$|\Psi_{nm}^{(lk)\prime}\rangle_{34} = N_n^{(lk)\prime}\left(A_{nm}^{(lk)}|01\rangle_{34} + |10\rangle_{34}\right), \qquad (51)$$

where $N_n^{(lk)\prime} = \left(1 + \left|A_n^{(lk)}\right|^2\right)^{-1/2}$ is a normalization factor. Here, modes 2 and 3 are mixed on balanced beam splitter (5a) with $t = r = 1/\sqrt{2}$ with subsequent registration of outcomes either $|01\rangle_{23}$ or $|10\rangle_{23}$ that leads to production of original unknown qubit with success probability

$$q_{nm}^{(lk)S} = \frac{\left|A_{nm}^{(lk)}\right|^2}{1+\left|A_{nm}^{(lk)}\right|^2}, \qquad (52)$$

where subscript $S$ concerns the fact that an unknown qubit was restored by the quantum swapping method. We note only the fact that amplitude demodulation by using amplitude displacement allows us to continue this procedure with the remaining states not satisfying the condition (49), while the quantum swapping procedure can only be performed once.

The same demodulation methods are applicable to the states (38). Then, we have the success probability for Bob to restore original unknown qubit from AM one

$$\delta P_T^{(lk)D} = F^2(\alpha)\sum_n^\infty \sum_p^\infty F^2(\gamma)|c_{ln}(\alpha)|^2 |c_{1p}(\gamma)|^2, \qquad (53)$$

where parameter $\gamma$ follows from relation

$$A_n^{(lk)}(\alpha)(c_{0n}(\gamma)/c_{1n}(\gamma)) = \pm 1. \qquad (54)$$

Another way to demodulate AM unknown qubits (38) allows for us to perform it with success probability

$$\delta P_T^{(lk)S} = F^2(\alpha)|c_{ln}(\alpha)|^2 \frac{\left|A_n^{(lk)}\right|^2}{1+\left|A_n^{(lk)}\right|^2}. \qquad (55)$$

We consider the case of $l < k$ and $n < m$. Let us start with the case of $l = 0$ and $k = 1$. Corresponding curves of $P_T^{(01)}$ and $P_{nm}^{(01)S}$ for different $n$ and $m$ in dependency on $\alpha$ are shown in the left part of the Fig. 2. Success probability to teleport unknown qubit without amplitude demodulation procedures takes maximal value $\left(P_T^{(01)}\right)_{max} = 0.2637$ under $\alpha = 0.628482$. The condition $A_{01}^{(01)} = A_{01}^{(01)} = -1$ is turned out to be satisfied in the case of $\alpha = 1/\sqrt{2}$. This allows us to increase the success probability $P_T^{(01)}(\alpha = 1/\sqrt{2}) = 0.2578$ by $0.18394$. Thus, the success probability for Alice to directly teleport to Bob unknown qubit becomes $P_S = P_T^{(01)}(\alpha = 1/\sqrt{2}) + P_{01}^{(01)S}(\alpha = 1/\sqrt{2}) = 0.441789$ as shown on the right side of the Fig. 2. At the same time, the probability $P_S = P_T^{(01)}(\alpha = 0.628482) + P_{01}^{(01)S}(\alpha = 0.628482) = 0.500673$ takes on greater value for the displacement amplitude corresponding to maximal value of $P_T^{(01)}$. But this probability consists of two events: the direct teleportation of an unknown qubit (without amplitude-distorting factor) and the teleportation with output AM qubit which needs an amplitude demodulation procedure. Consider the case of $l = 1$ and $k = 2$ whose functions $P_T^{(12)}$ and $P_{nm}^{(12)S}$ for different $n$ and $m$ in dependency on $\alpha$ are shown in the left part of the Fig. 3. The success probability $P_T^{(12)}$ has its maximum under $= 0.4072$ $\left(P_T^{(01)}\right)_{max} = 0.24371$. If we consider the contribution from the realization of the AM states with $A_{12}^{(12)}(\alpha = 0.4072) = A_{12}^{(12)-1}(\alpha = 0.4072)$, then this adds a value $P_{12}^{(12)S}(\alpha = 0.4072) = 0.2883$ to $\left(P_T^{(01)}\right)_{max}$, finally, resulting in $\left(P_T^{(01)}\right)_{max} + P_{12}^{(12)S}(\alpha = 0.4072) = 0.5317$ as shown on the right side of the Fig. 3. We have $A_{12}^{(12)}(\alpha =$ $0.5053) = A_{12}^{(12)-1}(\alpha = 0.5053) = -1$. Thus, choosing the value of $\alpha = 0.5053$, we get the probability of success of quantum teleportation of an unknown state (without amplitude-distorting factor) equal to $P_S = P_T^{(12)}(\alpha = 0.5053) + P_{12}^{(12)S}(\alpha = 0.5053) = 0.4014$.

Let us analyze the amplitude-distorting factors $A_{nm}^{(lk)}$. Two examples of the values of this parameter are given in Table 1 for $l = 0$ and $k = 1$ and Table 2 for $l = 1$ and $k = 2$, respectively.

| $n$ | 0 | 0 | 1 | 0 | 1 | 0 | 1 | 0 |
|---|---|---|---|---|---|---|---|---|
| $m$ | 2 | 3 | 2 | 4 | 3 | 5 | 4 | 5 |
| $A_{nm}^{(01)}$ | $-1/3$ | $-0.2$ | $1/3$ | $-1/7$ | $0.2$ | $-1/9$ | $1/7$ | $3/5$ |
| $A_{mn}^{(01)}$ | $-3$ | $-5$ | $3$ | $-7$ | $5$ | $-9$ | $7$ | $5/3$ |

**Table 1** Values of amplitude-distorting factors $A_{nm}^{(01)}(\alpha = 1/\sqrt{2})$ for different values of $n$ and $m$.

| $n$ | 0 | 0 | 0 | 0 | 1 |
|---|---|---|---|---|---|
| $m$ | 1 | 2 | 3 | 4 | 3 |
| $A_{nm}^{(12)}$ | $0.427$ | $-0.427$ | $-0.155$ | $-0.0954$ | $-0.362$ |
| $A_{mn}^{(12)}$ | $2.343$ | $-2.343$ | $-6.468$ | $-10.481$ | $-2.76$ |

**Table 2** Values of amplitude-distorting factors $A_{nm}^{(12)}(\alpha = 0.5053)$ for different values of $n$ and $m$.

Amplitude-distorting factors can be divided into two types: $A_{nm}^{(lk)} < 1$ and $A_{mn}^{(lk)} > 1$ provided that $n < m$. It follows from Eq. (52) the probability $q_{nm}^{(lk)S} \approx 1$ in the case of $A_{mn}^{(lk)} > 1$ that means quantum swapping procedure can be used to restore original unknown qubit from AM one with high probability. In the opposite case of AM state with amplitude-distorting factor $A_{nm}^{(lk)}$, the probability $q_{nm}^{(lk)S}$ takes small values. It turns out that the probability $F^4|c_{ln}|^2|c_{km}|^2$ can be much larger than $F^4|c_{lm}|^2|c_{kn}|^2$ i.e $F^4|c_{ln}|^2|c_{km}|^2 > F^4|c_{lm}|^2|c_{kn}|^2$. Then the main task is to search for demodulation procedure of the AM state with $A_{nm}^{(lk)} < 1$ which, for the time being, is quite a difficult problem. So, we have observed that overall success probability to teleport unknown qubit only using those two proposed demodulation methods becomes $P_T^{(01)O} = 0.522765$ and $P_T^{(12)O} = 0.4968$.

Similar difficulties occur in the demodulation of AM states (38) with amplitude-distorting factors $A_n^{(lk)}$ (Eq. (37)). Again, states with $A_n^{(lk)} > 1$ can be restored by quantum swapping procedure with probability (52) close to 1. The corresponding success probabilities $\delta P_T^{(lk)S}$ (Eq. (55)) and $\delta P_T^{(lk)D}$ (Eq. (53)) for teleporting and restoring AM unknown qubit depending on the displacement amplitude $\alpha$ are shown in Figure 4. It is worth noting Bob can continue the demodulation procedure in the case of use of method with displacement operator.

## 4. DV-CV quantum teleportation of unknown initially amplitude-distorting qubit

In the previous part, we showed the possibility for Bob to restore the original unknown qubit from the AM states with previously known amplitude-distorting factors $A_{nm}^{(lk)}$ and $A_n^{(lk)}$. These methods are probabilistic and allow us to demodulate the unknown qubit in the case of $A_{nm}^{(lk)} > 1$ and $A_n^{(lk)} > 1$ with high fidelity (52). In order to significantly increase the probability of success of the DV-CV quantum teleportation, we must increase the probability of demodulation of AM states with amplitude-distorting factors $A_{nm}^{(lk)} < 1$ and $A_n^{(lk)} < 1$.

Consider quantum teleportation of unknown qubit which was originally subjected to amplitude modulation by a third person, for example, Victor. The third party scheme is the most common. Victor prepares an unknown qubit and then checks the quality of the teleported qubit. Suppose, he prepares the following qubit

$$|\varphi_{AM}^{(01)}\rangle_{12} = N_{AM}^{(01)}\left(a_0|01\rangle_{12} + a_1 A_{01}^{(01)-1}|10\rangle_{12}\right), \tag{56}$$

with known amplitude-distorting factor $A_{01}^{(01)}$ and $a_0$, $a_1$ being the unknown amplitudes, where $N_{AM}^{(01)} = \left(1 + \left(\left|A_{01}^{(01)}\right|^{-2} - 1\right)|a_1|^2\right)^{-0.5}$ is a normalization factor. After preparing the AM qubit, Victor hands over it to Alice. The same entangled state (1) is used to implement DV-CV quantum teleportation of initially AM unknown qubit. Using the same mathematical apparatus, we can get similar expressions (19) but with different states $|\Psi_{nm}^{(lk)}\rangle$ (Eqs. (20),(21)). After Alice makes the parity measurement in the first mode and fixes $n$ and $m$ photons in the third and fourth modes, she can send information about them to Bob so that he can carry out unitary transformations (24,25) over his photon. Finally, Боб obtains the state

$$N_{nm}^{(01)\prime}\begin{bmatrix} a_0 \\ a_1 A_{01}^{(01)-1} A_{nm}^{(01)} \end{bmatrix}, \tag{57}$$

where $N_{nm}^{(01)\prime} = \left(1 + \left(\left|A_{01}^{(01)}\right|^{-2}\left|A_{nm}^{(01)}\right|^2 - 1\right)|a_1|^2\right)^{-0.5}$ is a normalization factor with probability

$$P_{nm}^{(01)} = \frac{F^4|c_{ln}(\alpha)|^2|c_{km}(\alpha)|^2 N_{AM}^{(01)2}}{N_{nm}^{(01)\prime 2}}. \tag{58}$$

In this case, the probability of success depends on the parameter $|a_1|$ of the unknown qubit due to the presence of members $N_{AM}^{(01)}$ and $N_{nm}^{(01)\prime}$ in formula (58).

The advantage of the initial modulation of an unknown qubit is that when fixing certain measurement outcomes, Bob gets the original unknown qubit with higher success probability. So if Alice registers the measurement outcomes $|01\rangle_{34}$, then Bob (after applying unitary transformations) receives the original unknown qubit as $A_{01}^{(01)-1}A_{01}^{(01)} = 1$ with the success probability

$$P_{01}^{(01)} = F^4|c_{ln}(\alpha)|^2|c_{km}(\alpha)|^2 N_{AM}^{(01)2}. \tag{59}$$

All other states resulting from the registration of other measurement outcomes contain an amplitude-distorting factor $A_{01}^{(01)-1}A_{nm}^{(01)}$. Bob can proceed to the demodulation procedure using the methods discussed above. So if he uses the quantum swapping method (Eqs. (51,52)) to get rid of amplitude-distorting factor, then, in the general case, the probability of success for Bob to get the original unknown quantum qubit becomes

$$P_T^{(01)} = F^4 N_{AM}^{(01)2} \begin{pmatrix} |c_{00}|^2|c_{11}|^2 + |c_{01}|^2|c_{10}|^2 \frac{\left|A_{10}^{(01)}\right|^4}{1+\left|A_{10}^{(01)}\right|^4} + \\ \frac{\left|A_{10}^{(01)}\right|^2}{1+\left|A_{10}^{(01)}\right|^2}\sum_{n=0}^{\infty}|c_{0n}|^2|c_{1n}|^2 + \\ \sum_{n}^{\infty}\sum_{m,n\neq m,n+m>1}^{\infty}|c_{0n}|^2|c_{1m}|^2 \frac{\left|A_{10}^{(01)}\right|^{-2}\left|A_{nm}^{(01)}\right|^2}{1+\left|A_{10}^{(01)}\right|^{-2}\left|A_{nm}^{(01)}\right|^2} \end{pmatrix}. \tag{60}$$

The contribution of only a few events in is significant. The contribution of the overwhelming number of events in (60) is very small and can be neglected. The corresponding plots of the success probability $P_T^{(01)}$ depending on the parameter $|a_1|$ of unknown qubit are shown in Figure 5 (left side of the figure) for different values of the displacement amplitude $\alpha$. As can be seen from these plots, there are values of $|a_1| \ll 1$ for which the probability of success can take values close to one. Thus, the method of initial amplitude modulation of an unknown qubit can lead to an increase in the efficiency of the DV-CV quantum teleportation.

Consider another possibility to implement DV-CV quantum teleportation of initially AM unknown qubit. Suppose Victor prepares the next unknown qubit

$$|\varphi_{AM}^{(01)}\rangle_{12} = N_{AM}^{(01)}\left(a_0|0\rangle_1 + a_1 A_0^{(01)-1}|1\rangle_1\right), \tag{61}$$

where $N_{AM}^{(01)} = \left(1 + \left(\left|A_0^{(01)}\right|^{-2} - 1\right)|a_1|^2\right)^{-0.5}$ is a normalization factor, and transmit it to Alice. Amplitudes $a_0$ and $a_1$ of the state (61) are unknown to anyone, while amplitude-distorting factor $A_0^{(01)-1}$ follows from (37). Alice's unknown AM qubit interacts with an entangled hybrid state (1) on HTBS as shown in Figure 1. After Alice performs the measurement in the auxiliary modes (36) and sends the measurement results to Bob on the classical channel, he can implement the corresponding unitary transformations on his dual-rail single photon. The result of this procedure is the following state

$$N_n^{(01)\prime}\begin{bmatrix} a_0 \\ a_1 A_0^{(01)-1} A_n^{(01)} \end{bmatrix}, \tag{62}$$

where $N_n^{(01)\prime} = \left(1 + \left(\left|A_0^{(01)}\right|^{-2}\left|A_n^{(01)}\right|^2 - 1\right)|a_1|^2\right)^{-0.5}$ is the normalization factor of obtained state. Success probability to get the state is

$$P_n^{(01)} = \frac{F^2|c_{0n}(\alpha)|^2 N_{AM}^{(01)2}}{N_n^{(01)\prime 2}}. \tag{63}$$

The probability of success includes the normalization parameters $N_{AM}^{(01)}$ and $N_n^{(01)\prime}$, so it depends on the parameter $|a_1|$ of the unknown qubit. If Alice registered the vacuum in the auxiliary second mode, then Bob has the initial unknown qubit (after the implementation of the corresponding unitary transformations), since $A_0^{(01)-1}A_0^{(01)} = 1$. Success probability for Alice to register such outcome becomes

$$P_0^{(01)} = F^2|c_{0n}(\alpha)|^2 N_{AM}^{(01)2}, \tag{64}$$

as $N_0^{(01)\prime} = 1$. If Alice registers a non-vacuum outcome $|n\rangle$ with $n \ne 0$, then Bob's state contains amplitude-distorting factor $A_0^{(01)-1}A_n^{(01)}$. In the case, Bob can use one of the two considered methods for demodulating the AM states with corresponding success probabilities. Consider the method of demodulation of the AM states using a coherent state of large amplitude (Eq. (48)). To use this method, one needs to find the value of the parameter $\gamma$ (Eq. (54)) which greatly complicates the analytical view of the probability of success to teleport unknown qubit and get rid of the amplitude-distorting factor. The corresponding dependences of the success probability $P_T^{(01)}$ of the initially AM unknown qubit depending on the parameter $|a_1|$ of the unknown qubit for different values of the displacement amplitudes $\alpha$ are shown in the right part of Figure 5. As can be seen from these graphs, the probability of success can be significantly increased compared to the case discussed above.

## 5. Results

We considered the ability to teleport an unknown qubit using DV-CV interaction mechanism. This mechanism is implemented in the interaction of CV and DV states on HTBS. Non-maximally entangled hybrid state composed of coherent components with opposite in sign amplitudes and DV sate is used to perform DV-CV quantum teleportation of unknown qubit. The coherent components of the state (1) displace the unknown state to equal modulo but opposite in sign amplitudes in an indistinguishable manner so that all information about the value of the displacement disappears. The unknown state can be displaced by both positive and negative values. If an unknown qubit is displaced by a positive value, then the relative phase of the decomposition coefficients of the displaced states in the measurement basis does not change regardless of the parity of the basic states. On the contrary, the relative phase of the displaced unknown qubit in the measurement basis can vary on the opposite depending on the parity of the basis states in the case of the negative amplitude of displacement. This nonlinear effect akin to the action of controlled−$Z$ gate is a base of DV-CV quantum teleportation of unknown qubit. Bob, having received a limited amount of classical information about Alice's results of the measurements, can perform the appropriate set of unitary transformations over his single photon. Since the amplitudes of the decomposition of the displaced states of light in the measuring basis are not equal to each other, the teleported states acquire additional known amplitude-distorting coefficients. The presence of an amplitude-distorting factor in the teleported qubit can be recognized as an inherent trait of the DV-CV quantum teleportation. It may recall the CV deterministic quantum teleportation of an unknown qubit whose output fidelity suffers due to the absence of the maximally entangled quantum channel. On the contrary, DV teleportation allows us to get output state with a high degree of fidelity (in the ideal case with unit fidelity), but the implementation of the full Bell-states measurement using linear optics is impossible, which reduces the probability of success up to 0.5. All measurement outcomes in DV-CV teleportation are distinguishable. But the fidelity of the output state in Bob's hands is also not ideal as in CV teleportation. Although it is worth noting that the protocol participants know the values of amplitude-distorting factors. Notice the difference of quantum channels used in CV and DV-CV quantum teleportation. Two-mode squeezed vacuum is used in CV teleportation. The description and predictions of the protocol are based on quantum nonlocality of entangled quantum state (1) and cannot be simulated by any local hidden variable theory.

The key issue for increasing the protocol's efficiency is resolving the demodulation problem or the same getting rid of an unknown quantum state from a previously known amplitude-distorting factor, preferably with a success rate close to unity. We considered only two such probabilistic possibilities which are based on the displacement of the state in the auxiliary mode with the subsequent registration of some events in the measurement number state basis and the quantum swapping procedure. Moreover, it is worth noting that this mechanism can be implemented in various interpretations, some of which we have considered. Each of the considered schemes allows us to calculate the amplitude-distorting factors (Eqs. (16,37)). So, the optical scheme shown in Figure 1 with the additional interaction of an unknown qubit with coherent state allows us to teleport it without demodulation procedures if detectors register the same number of photons in the auxiliary modes. Other interpretations that could increase the efficiency of the DV-CV quantum teleportation are possible. So, we did not consider amplitude-distorting factors (16) with different displacement amplitudes $\alpha \neq \alpha_1$. We used the hybrid state (1) with dual-rail single-photon at Bob's location. In fact, the same universal mechanism works if we use a quantum state (1) with different state in Bob's hands, including states from other physical systems, which could increase the success probability of the demodulation procedure. Consideration of these issues requires separate investigation. Within the considered interpretations of the DV-CV quantum teleportation of unknown qubit $|\varphi^{(lk)}\rangle$ (Eq. (3)) with small values of $l$ and $k$, it is necessary to recognize that the scheme with the initial amplitude modulation of the unknown qubit is the most effective (Fig. 5).

**Acknowledgement**

The work was supported by Act 211 Government of the Russian Federation, contract № 02.A03.21.0011.

**List of figures**

**Figure 1**
A schematic representation of implementation of DV-CV quantum teleportation with help of hybrid non-maximally entangled state (1). Coherent components interact with unknown qubit in an indistinguishable manner on HTBS. Measurements made at a microscopic level allows for Bob to obtain (after the corresponding unitary transformations initiated by the classical information (CI) from Alice) set of the states depending on Alice's measurement outcomes due to quantum nonlocality. Part of states is the original unknown states, while the others acquire additional amplitude known factors. DV-CV quantum teleportation can be performed in various interpretations in order to influence which part of the teleported qubits is original unknown state and which is AM states. Different implementation schemes also determine the amplitude-distorting factors of the output states. So, if the scheme involves additional HTBS for interaction of coherent state $|0, -\beta\rangle_2$ with the original state (4), then we deal with amplitude-distorting factors (16). If the scheme without additional HTBS is used, then participants work with amplitude-distorting factors (37). Another interpretation includes the third party that initially generates AM unknown qubit. The considered schemes should also include a demodulation procedure (DP) in order to get rid of amplitude-distorting factors. Commercially achievable avalanche photodiode (APD) being a highly sensitive semiconductor electronic device are used for registration of the measurement outcomes. Photon number resolving detector is used in first (coherent) mode to determine the parity of the SCS. $S$ means a source of the hybrid entangled state (61).

**Figure 2**
Plots of the success probabilities $P_T^{(01)}$ and $P_{nm}^{(01)}$ for different $n$ and $m$ in dependency on the displacement amplitude $\alpha$ (in the left side of the picture). Only three graphs of probabilities $P_T^{(01)}$, $P_{01}^{(01)S}$, giving the maximum contribution, and $P_S = P_T^{(01)} + P_{01}^{(01)S}$ are left on the right side of the figure. The value $P_S = 0.4418$ corresponds to quantum teleportation of unknown qubit without amplitude-distorting factor.

**Figure 3**
Plots of the success probabilities $P_T^{(12)}$ and $P_{nm}^{(12)}$ for different $n$ and $m$ in dependency on the displacement amplitude $\alpha$ (in the left side of the picture). Only three graphs of probabilities $P_T^{(12)}$, $P_{12}^{(12)S}$, giving the maximum contribution, and $P_S = P_T^{(12)} + P_{12}^{(12)S}$ are left on the right side of the figure. The value $P_S = 0.401$ corresponds to quantum teleportation of unknown qubit without amplitude-distorting factor.

**Figure 4**
Plots of the success probabilities $\delta P_T^{(01)S}$ and $\delta P_T^{(01)D}$ to teleport and restore (get rid of amplitude-distorting factor by one two proposed methods) unknown qubit in dependency on the displacement amplitude $\alpha$.

**Figure 5**
Plots of the success probabilities $P_T^{(01)}$ to teleport and restore (get rid of amplitude-distorting factor by one two proposed methods) unknown initially AM qubit in dependency on $|a_1|$ of unknown qubit for the different values of the displacement amplitude $\alpha$. The left side shows the success probability when using the initially modulated unknown qubit (56) with amplitude-distorting factor $A_{01}^{(01)-1}$. The quantum swapping method (Eqs. (51,52)) is used to get rid of amplitude-distorting factors (Eq. (60)). The right side of the graph shows the success probability of teleporting AM unknown qubit (61), where original state is restored with help of mixing AM unknown qubit with coherent state (Eqs. (48,49)).

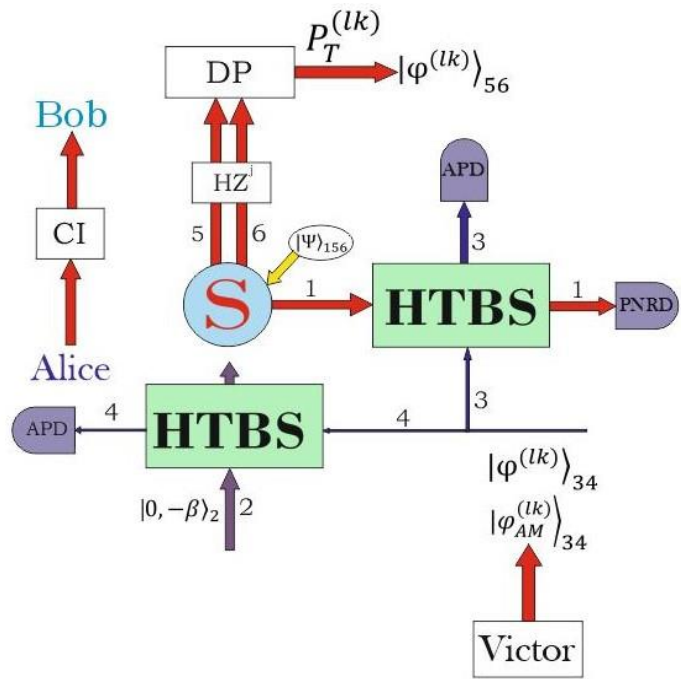

**Fig. 1**

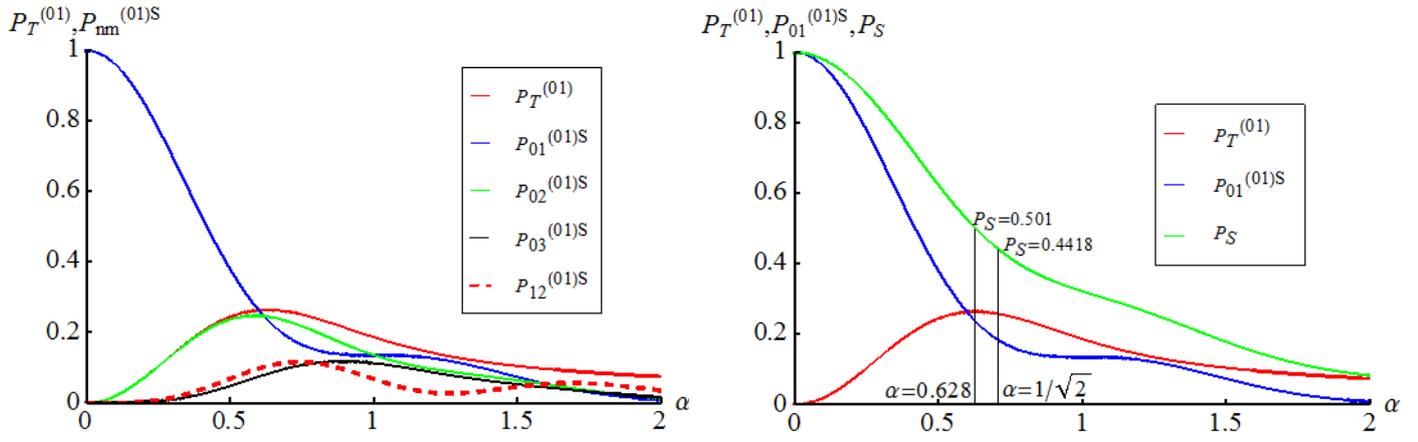

**Fig. 2**

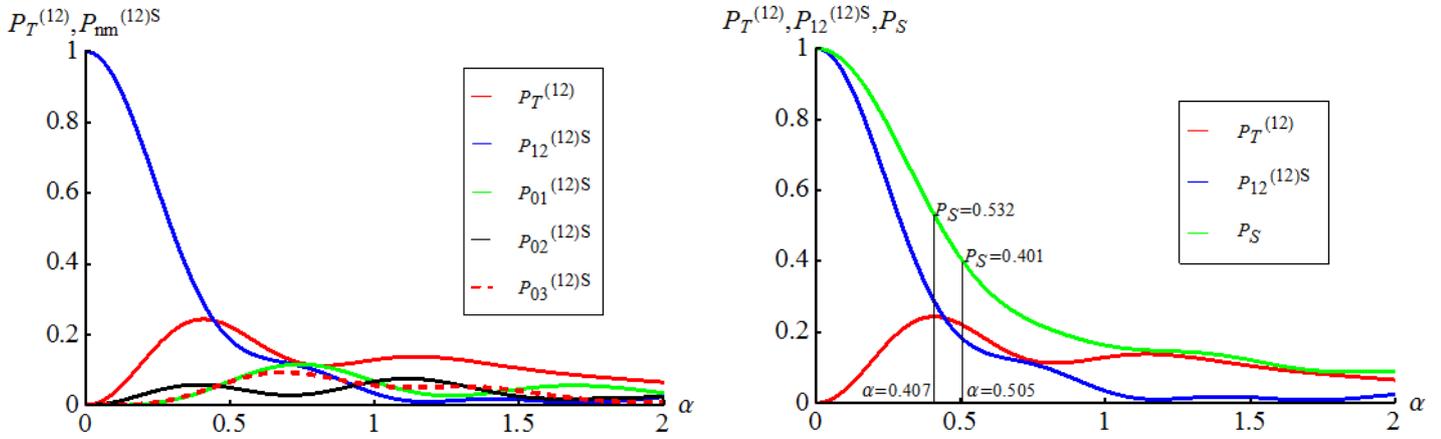

**Fig. 3**

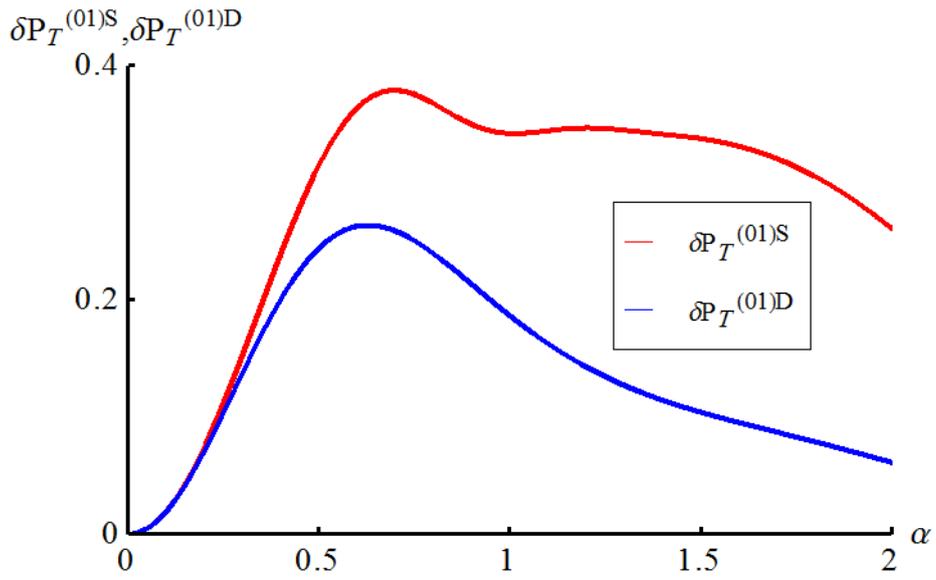

**Fig. 4**

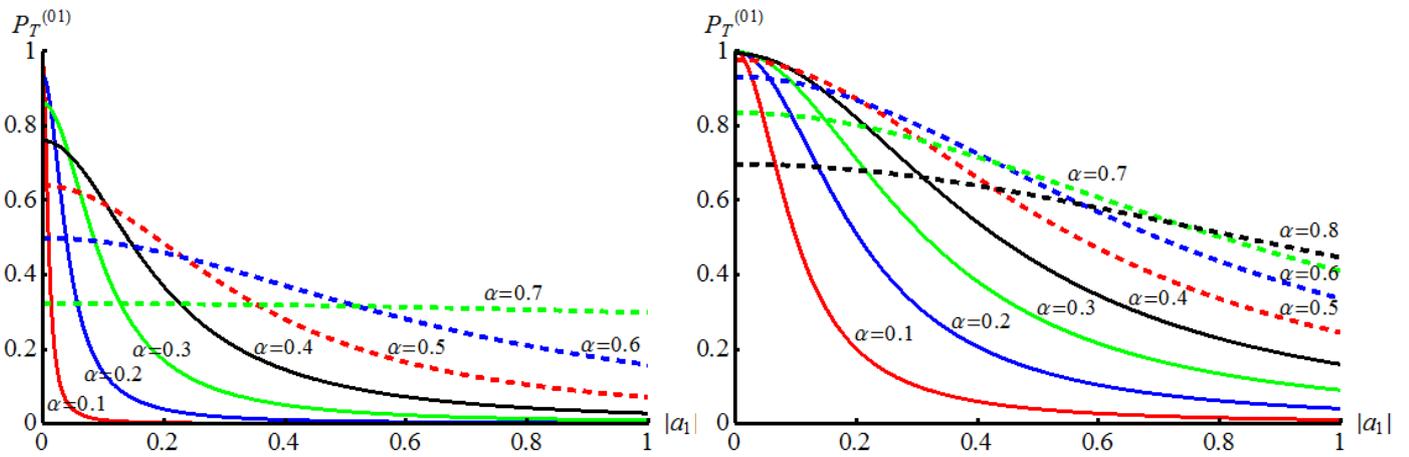

**Fig. 5**